\documentclass{rspublic}
\usepackage{graphicx}
\usepackage{epstopdf}
\begin{document}
\title[Kagome spin ice]{Magnetic charge and ordering in kagome spin ice}
\author[G.-W. Chern and O. Tchernyshyov]{Gia-Wei Chern$^{1,2}$ and Oleg Tchernyshyov$^3$}
\affiliation{$^1$Department of Physics, University of Wisconsin, Madison, WI 53706, USA \\
$^2$Institute for Complex Adaptive Matter, University of California, Davis, CA 95616 USA\\
$^3$Department of Physics and Astronomy, Johns Hopkins University, Baltimore, MD 21218, USA}
\maketitle

\begin{abstract}{Spin ice, magnetic ordering, magnetic charge}
We present a numerical study of magnetic ordering in spin ice on kagome, a two-dimensional lattice of corner-sharing triangles. The magnet has six ground states and the ordering occurs in two stages, as one might expect for a six-state clock model. In spin ice with short-range interactions up to second neighbors, there is an intermediate critical phase separated from the paramagnetic and ordered phases by Kosterlitz-Thouless transitions. In dipolar spin ice, the intermediate phase has long-range order of staggered magnetic charges. The high and low-temperature phase transitions are of the Ising and 3-state Potts universality classes, respectively. Freeze-out of defects in the charge order produces a very large spin correlation length in the intermediate phase. As a result of that, the lower-temperature transition appears to be of the Kosterlitz-Thouless type.
\end{abstract}

\section{Introduction}

Spin ice is a ferromagnet with peculiar properties (Gingras 2011).  Its discovery in the rare-earth pyrochlore Ho$_2$Ti$_2$O$_7$ by Harris \emph{et al.} (1997) attracted interest of physicists because geometric frustration results in an enormous number of ground states in spin ice. The degeneracy is similar to that of proton positions in water ice (Petrenko \& Whitworth 1999) and manifests itself in a large residual entropy at low temperatures measured by Ramirez \emph{et al.} (1999). Later it was shown that spin ice has another peculiar feature: even though it remains disordered as the temperature goes to zero, spin correlations decay with the distance algebraically rather than exponentially (Isakov \emph{et al.} 2004; Henley 2005). Such critical behavior is characteristic of systems with constraints exemplified by lattice models with hardcore dimers. In the case of spin ice, at low temperatures each tetrahedron of rare-earth ions is forced to have two spins pointing toward the center of the tetrahedron and two away from it.

The current resurgence of interest in spin ice stems from an unusual character of magnetic excitations in this class of materials (Ryzhkin 2005; Castelnovo, Moessner \& Sondhi 2008). Reversing a single spin in a spin-ice state violates the above-mentioned constraint on the two tetrahedra sharing that spin. One of them has three spins pointing in and one pointing out, the other one spin pointing in and three out. By reversing additional spins, the two defects can be moved around independently from each other. These two defect tetrahedra carry equal and opposite magnetic charges defined as sources and sinks of magnetic field $\mathbf H$. A defect with magnetic charge $Q$ in an external magnetic field $\mathbf H_\mathrm{ext}$ experiences a Zeeman force $\mu_0 Q \mathbf H_\mathrm{ext}$.  Two defects with magnetic charges $Q_1$ and $Q_2$ interact with one another via a Coulomb potential $\mu_0 Q_1 Q_2/(4 \pi r)$. Various properties of spin ice can be most naturally described by focusing on the motion of these defects (Fennell \emph{et al.} 2009; Jaubert \emph{et al.} 2009; Morris \emph{et al.} 2009; Ryzhkin \& Ryzhkin 2011). 

The concept of magnetic charge is not exactly novel. It can be found in earlier research articles (Saitoh \emph{et al.} 2004) and even in textbooks (Landau \& Lifshitz 1984). Magnetic charges in spin ice are remarkable because they are mobile and represent a rare example of fractionalized excitations in three spatial dimensions: the underlying degrees of freedom, magnetic dipoles, must be split in half, so to speak, to create magnetic monopoles. 

It is worth pointing out that magnetic charges in spin ice are fundamentally different from the magnetic monopoles introduced by Dirac (1931) to explain quantization of electric charge. They are sources and sinks of magnetic field $\mathbf H$, rather than of magnetic induction $\mathbf B$. For that reason, there are no restrictions on the possible values of magnetic charges in spin ice. 

\begin{figure}
\includegraphics[width=\columnwidth]{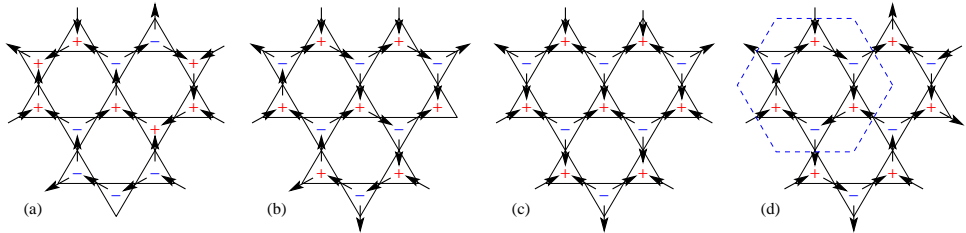}
\caption{(a) Generic spin-ice state on kagome. (b) A spin-ice (micro)state with magnetic charges ordered in a staggered patter. (c) Expectation values of spins and magnetic charges in a charged-ordered intermediate state. Note that the expectation values for the Ising variables are $\langle \sigma_i \rangle = \pm 1/3$ in this state, so the charges on triangles are $\langle Q_\alpha \rangle  =\pm 1$. (d) A ground state of kagome spin ice with ordered charges and magnetic dipoles. The dashed line is the perimeter of a magnetic unit cell.}
\label{fig:kagome}
\end{figure}

In this paper, we illustrate the utility of the concept of magnetic charge for the problem of spin ice on a different lattice. A natural generalization of the pyrochlore network is a lattice consisting of corner-sharing simplexes (Chalker 2011). In two dimensions, corner-sharing triangles form kagome, Fig.~\ref{fig:kagome}. A spin is shared by two triangles and points along the line connecting their centers. The easy axes of the three spins on a triangle make angles of $120^\circ$ with each other. Like on the pyrochlore lattice, interactions of the antiferromagnetic kind are not frustrating: an isolated triangle has two ground states with all spins pointing in or all pointing out. A kagome lattice with nearest-neighbor antiferromagnetic interactions also has two ground states, in which spins point into triangles of the same orientation. On the other hand, ferromagnetic interactions are frustrating and yield six ground states on a single triangle, with two spins pointing in (and one out) or two pointing out (and one in). The number of ground states grows exponentially with the number of spins when interactions are ferromagnetic and only include nearest neighbors. In that, spin ice on kagome is quite similar to its analog on the pyrochlore lattice. However, the addition of longer-range interactions, especially dipolar ones, reveals some important differences between the two systems. 

The main difference between the pyrochlore network and kagome concerns the allowed values of magnetic charge $Q$ of a simplex defined as the flux of magnetization $\mathbf M$ into the simplex. For a tetrahedron, $Q$ is even (in the appropriate units): zero in a two-in-two-out state, $\pm 2$ in three-in-one-out or three-out-one-in states and $\pm 4$ in the all-in or all-out states. The energy of spin ice penalizes states with a large (in absolute value) charge $Q$ so that simplexes are neutral, $Q=0$, in low-energy states. The lack of magnetic charges on tetrahedra explains a puzzling observation by Gingras and coworkers: spin-ice states remain very nearly degenerate even in the presence of long-range dipolar interactions (Gingras \& den Hertog 2001). As Castelnovo \emph{et al.} (2008) showed, the energy of dipolar interactions is well approximated by the Coulomb energy of magnetic charges residing on simplexes. Because these charges vanish in two-in-two-out states, their Coulomb energy is the same. The residual interactions, neglected in the Coulomb approximation, force the system into a phase with long-range magnetic order at temperatures low compared to the strength of dipolar interactions (Melko \emph{et al.} 2001).

In contrast, the magnetic charge of a triangle is $\pm 1$ in a spin-ice state and $\pm 3$ when its spins point all in or all out. Even if the interactions tend to suppress magnetic charges, the lowest (in magnitude) values of charge on a triangle are $\pm 1$. This has important consequences that we briefly outline below and consider in detail in the rest of the paper.

First, the presence of a charge degree of freedom on simplexes makes spin ice on kagome much more fluid than its pyrochlore counterpart. In pyrochlore spin ice, spin dynamics slows down considerably at low temperatures. Single-spin flips result in the creation of two magnetic charges and thus become forbidden when the temperature falls below the energy cost of a monopole. Moves within the low-energy sector require flipping entire chains of alternating spins with a minimal length of six (Melko \emph{et al.} 2001). Alternatively, spin fluctuations can be mediated by the motion of a few remaining magnetic monopoles. Experimental studies confirm a very slow relaxation in spin ice at liquid-helium temperatures (Snyder \emph{et al.} 2001; Giblin \emph{et al.} 2011), although the primary mechanism of low-$T$ spin dynamics in the rare-earth pyrochlores remains to be clarified. No such impediment to spin fluctuations exists in kagome ice because a single spin flip does not necessarily take the magnet out of the low-energy sector. Such a local move is allowed if the magnetic charges of the two simplexes sharing the spin change from $+1$ and $-1$ to $-1$ and $+1$. Although at present there are no materials realizing two-dimensional kagome spin ice, several experimental groups have made artificial magnetic arrays with this geometry (Tanaka \emph{et al.} 2006). Qi \emph{et al.} (2008) demonstrated that their artificial spin ice on kagome stayed strictly within the low-energy sector in which the magnetic charges remain $\pm 1$. Ladak \emph{et al.} (2010) observed triple charges, but this is likely the result of strong quenched disorder in their samples (Ladak \emph{et al.} 2011). 

Second, the presence of a charge degree of freedom opens a possibility of new kinds of ordered phases in spin ice. Pyrochlore spin ice is expected to have only two thermodynamics phases. As the temperature is lowered from $+\infty$, the paramagnetic state gradually crosses over to the spin-ice state, which still preserves all of the symmetries of the Hamiltonian and thus remains a paramagnet, albeit a correlated one, from the standpoint of the Landau theory. At a sufficiently low temperature, the residual spin interactions not captured by the Coulomb model (Castelnovo \emph{et al.} 2008) induce a phase transition into an ordered state that breaks the time-reversal and lattice symmetries (Melko \& Gingras 2004). On kagome ice, one has good reasons to expect an intermediate ordered phase with staggered magnetic charges that minimizes the Coulomb energy (M{\"o}ller and Moessner 2009; Chern \emph{et al.} 2011). This phase is characterized by very strong spin fluctuations. At the lowest temperatures, the symmetry is broken further and the system settles into a state with long-range magnetic order. 

To test these heuristic considerations, we have performed extensive numerical studies of spin ice on kagome, both with short-range interactions (nearest and next-nearest neighbors) and with long-range dipolar interactions. The problem turned out to be subtle and the answers different for the models with short and long-range interactions. In both cases, there is an intermediate phase. However, the nature of the intermediate phase is drastically different. The dipolar spin ice has the expected charge-ordered phase, whereas the short-range model has a critical intermediate phase without true long-range order. The determination of the universality classes of the phase transitions required rather large system sizes because some of the transitions were of the Kosterlitz-Thouless type. Simulating large systems is particularly difficult in the presence of long-range interactions.

\section{Kagome spin ice: the model}
\label{sec:model}

\subsection{The Hamiltonians}

The model of spin ice on kagome was first introduced by Wills \emph{et al.} (2002). It has Ising spins $\mathbf S_i = \sigma_i \hat {\mathbf e}_i$ living on the vertices of corner-sharing triangles of a kagome lattice. The unit vector $\hat {\mathbf e}_i$ points along the line connecting the centers of the two triangles, while the Ising variable $\sigma_i$ encodes the state of the spin. It is convenient to choose the unit vectors $\hat {\mathbf e}_i$ in such a way that they point into triangles of one orientation (say, $\triangle$). The Hamiltonian of the model with nearest-neighbor (nn) exchange $J_1$ can be written in two ways: 
\begin{equation}
H_1 = -J_1 \sum_{\mathrm{nn}} \mathbf S_{i} \cdot \mathbf S_j 
= \frac{J_1}{2} \sum_{\mathrm{nn}} \sigma_{i} \sigma_j,
\end{equation}
where we relied on the result $\hat {\mathbf e}_i \cdot \hat {\mathbf e}_j = -1/2$ for nearest neighbors. As usual, a ferromagnetic exchange $J_1 > 0$ for spins $\mathbf S_i$ translates into an antiferromagnetic interaction for the Ising variables $\sigma_i$. It is well known that the Ising antiferromagnet on kagome remains paramagnetic down to zero temperature (Sy{\^o}zi 1951). It retains a finite entropy density in the ground state, 0.502 per spin (Kano \& Naya 1953). 

To understand this residual spin degeneracy, we rewrite $H_1$ in terms of magnetic charges of triangles
\begin{equation}
	\label{eq:H1b}
	H_1 = \frac{J_1}{4} \sum_{\alpha} Q_{\alpha}^2.
\end{equation}
Here we have neglected an irrelevant constant. The magnetic charge is defined as
\begin{equation}
	Q_{\alpha} = \pm \sum_{i\in \alpha} \sigma_i,
\end{equation}
with the plus sign for one type of triangles and minus for the other. As discussed above,
the allowed values of magnetic charges are $Q_\alpha = \pm 3$ and $\pm 1$. Eq.~(\ref{eq:H1b}) shows that 
triangles with triple charges are energetically disfavored. Consequently, as temperature is lowered the magent
gradually enters the spin-ice phase consisting exclusively of triangles with $Q_\alpha = \pm 1$, corresponding to
the two-in-one-out and one-in-two-out ice rules on kagome. 
The number of the spin-ice microstates grows exponentially with the number of spins. Invoking the famous 
Pauling estimation gives an entropy density $S_{\rm ice} = (1/3)\ln (9/2) = 0.5014$ per spin, which agrees 
very well with the numerical results.

Wills \emph{et al.} (2002) added interactions between next-nearest neighbors (nnn): 
\begin{equation}
H_2 = -J_2 \sum_{\mathrm{nnn}} \mathbf S_{i} \cdot \mathbf S_j 
= \frac{J_2}{2} \sum_{\mathrm{nnn}} \sigma_{i} \sigma_j.
\end{equation}
Whereas they considered both signs of $J_2$, we are interested in the antiferromagnetic case $J_2 < 0$.  M{\"o}ller and Moessner (2009) and our group (Chern \emph{et al.} 2011) added long-range dipolar interactions,
\begin{equation}
H_\mathrm{d} = \frac{Dr_\mathrm{nn}^3}{2} \sum_{i \neq j} \sigma_i \sigma_j
    \frac{(\hat {\mathbf e}_i \cdot \hat {\mathbf e}_j) - 3(\hat {\mathbf e}_i \cdot \hat {\mathbf r}_{ij})(\hat {\mathbf e}_j \cdot \hat {\mathbf r}_{ij})}{|{\mathbf r}_i - {\mathbf r}_j|^3},
\end{equation}
where $D = \mu_0 \mu^2/(4\pi r_\mathrm{nn}^3)$ is the characteristic
strength of dipolar coupling, $\mathbf r_i$ are spin locations,
$\hat {\mathbf r}_{ij} = ({\mathbf r}_i - {\mathbf r}_j)/|{\mathbf r}_i -
{\mathbf r}_j|$, $\mu$ is the magnetic dipolar moment of a spin and $r_\mathrm{nn}$ is the distance between nearest
neighbors. For brevity, we shall refer to the model of Wills \emph{et al.} (2002), with the Hamiltonian $H_1 + H_2$ and antiferromagnetic $J_2<0$, as the \emph{short-range kagome ice} and call the model with dipolar interactions, with the Hamiltonian $H_1 + H_\mathrm{d}$, the \emph{long-range kagome ice}. 

\subsection{The ground states}

\begin{figure}
\includegraphics[width=\columnwidth]{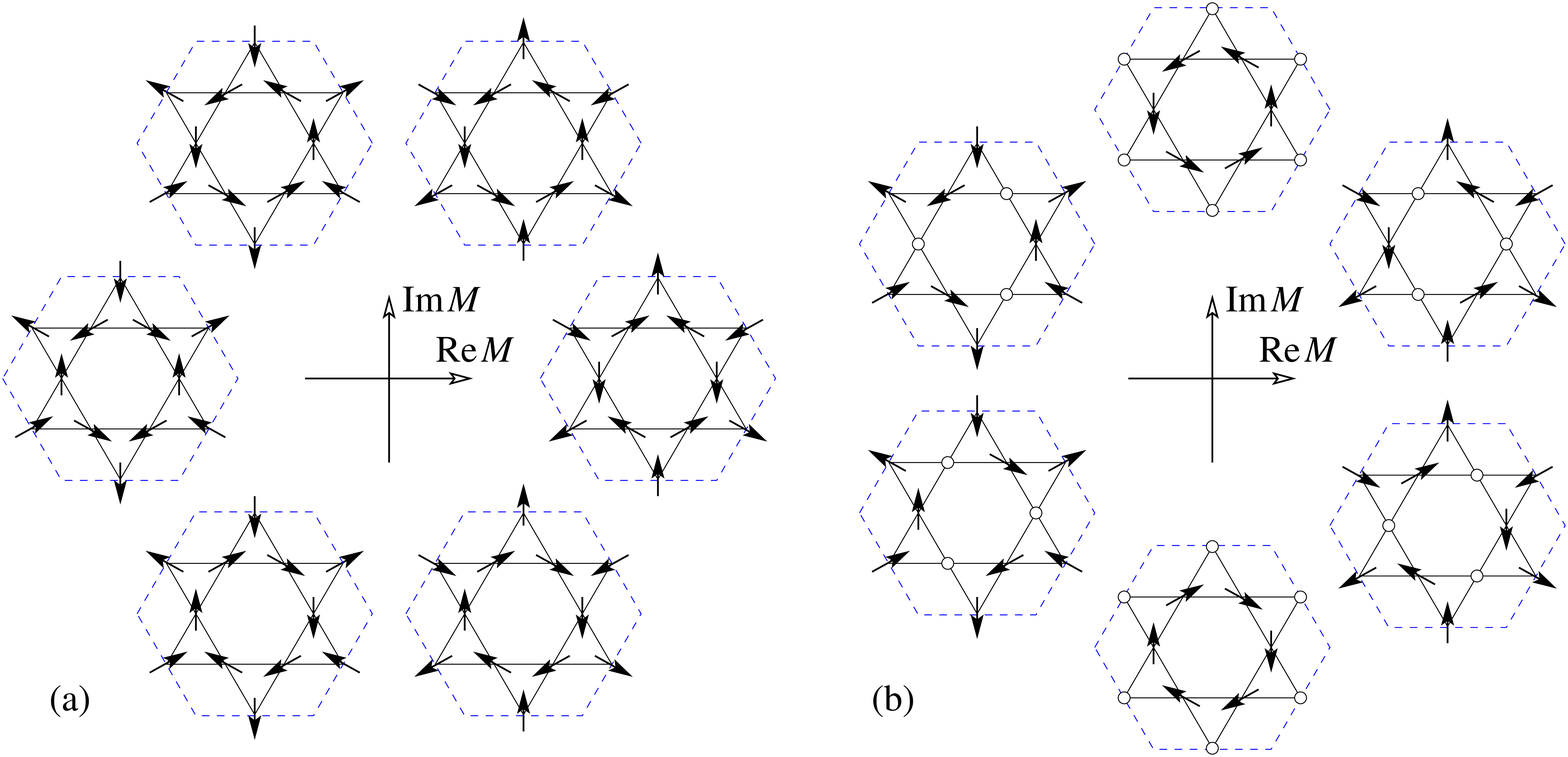}
\caption{(a) The ground states of the dipolar spin ice on kagome have a complex order parameter $M = |M|\re^{\ri\phi}$ with $\phi = n\pi/3$. (b) A proposed intermediate phase (Takagi \& Mekata 1993; Wills \emph{et al.} 2002) has $\phi = (n+1/2)\pi/3$.  In both cases, the magnetic unit cell contains 9 sites.}
\label{fig:clock}
\end{figure}

In both models, interactions between Ising variables $\sigma_i$ are antiferromagnetic for nearest neighbors and ferromagnetic for next-nearest neighbors. On the basis of that, one might expect the same type of magnetic order at low temperatures.  The ground states for the Ising model on kagome with first and second-neighbor interactions are known exactly (Wolf \& Schotte 1988; Takagi \& Mekata 1993). They have an enlarged ($\sqrt{3} \times \sqrt{3}$) magnetic unit cell with 9 spins. The corresponding ground states of the ice model are shown in Fig.~\ref{fig:clock}(a). With the aid of a complex order parameter, 
\begin{equation}
M = \frac{1}{N}\sum_{i} \sigma_i \exp{\left(\ri \mathbf Q \cdot {\mathbf r}_i\right)},
\label{eq:M-def}
\end{equation}
where $\mathbf Q = (4\pi/3,0)$ and $N$ is the number of spins, we can see that they are similar to those of the six-state clock model.  The phase of the order parameter $M = |M|\re^{\ri \phi}$ takes on the six values $\phi = n \pi/3$, where $n$ is an integer, Fig.~\ref{fig:clock}(a).

We were able to show that the long-range model with dipolar interactions has the same ground states. To do so, we treated the Hamiltonian $H_1+H_\mathrm{d} = \sigma_i A_{ij} \sigma_j/2$ as a quadratic form in the Ising variables. We replaced the constraint $\sigma_i = \pm 1$ with a less stringent normalization condition, $\sigma_i \sigma_i = N$, where $N$ is the number of sites, and minimized the quadratic form by finding the lowest (most negative) eigenvalue of the matrix $A_{ij}$. Although this procedure minimizes the energy over an enlarged space of states, the method gives the right answer if the eigenstate with the lowest eigenvalue has $\sigma_i = \pm 1$. That is indeed the case for the Hamiltonian $H_1+H_\mathrm{d}$. The corresponding eigenstates are the six ground states of the Ising model on kagome. 

\subsection{Phase transitions: general considerations}
\label{sec:transitions}

Previous theoretical studies have revealed that the six-state clock model in two dimensions may order in a number of scenarios (Cardy 1980). In all of them, there is a high-temperature disordered phase with $\langle M \rangle = 0$ and a low-temperature ordered phase with $\langle M \rangle \neq 0$. 

\begin{itemize}

\item[(A)] Two Kosterlitz-Thouless (KT) transitions with an intermediate critical phase in which $\langle M \rangle = 0$ and spin correlations decay as a power of the inverse distance. 

\item[(B)] From the paramagnetic phase, the system enters a partially ordered phase via an Ising transition. A second transition of the 3-state Potts universality class takes the system to the ordered state. The intermediate phase has an Ising order parameter $\langle M^3 \rangle \neq 0$, whereas $\langle M \rangle = 0$.

\item[(C)] Similar to (B) but with the Ising and Potts transitions exchanged. The intermediate phase has a 3-state Potts order parameter $\langle M^2 \rangle \neq 0$, whereas $\langle M \rangle = 0$.

\item[(D)] A discontinuous phase transition between the paramagnetic and fully ordered phase. 

\end{itemize}

Numerical studies (Wolfe \& Shotte 1988; Takagi \& Mekata 1993) provide compelling evidence for scenario (A) with two KT transitions in short-range Ising models on kagome. Takagi \& Mekata mistakenly ascribed a nonzero order parameter $\langle M \rangle$ to the intermediate phase, which is expected to have quasi-long-range order with power-law spin correlations and $\langle M \rangle$ = 0. 

Wills \emph{et al.} (2002) also suggested an intermediate ordered phase with six possible states shown in Fig.~\ref{fig:clock}(b). They are equivalent to the intermediate states of Takagi \& Mekata and have a nonzero order parameter $\langle M \rangle$ with $\phi = (n+1/2)\pi/3$. Such an intermediate state is clearly inconsistent with any of the four scenarios mentioned above. 

The intermediate critical phase in scenario (A) is most simply understood in the continuous version of the clock model, namely the XY model with a sixfold anisotropy term $a_6\cos{6\phi}$ in the free-energy functional. This term is irrelevant in the renormalization-group sense above the lower phase transition, so that the intermediate phase behaves as if the sixfold anisotropy were absent. In principle, there is a possibility that the sixfold anisotropy changes sign: at low temperatures, $a_6 < 0$ yields the states shown in Fig.~\ref{fig:clock}(a), whereas at higher temperatures, $a_6 > 0$ stabilizes the states shown in Fig.~\ref{fig:clock}(b). However, if the intermediate phase is indeed ordered and fully breaks the $Z_6$ symmetry, the transition between that phase and the fully symmetric paramagnet should follow one of the above scenarios, which requires yet another intermediate phase or a discontinuous phase transition. We see no evidence for either possibility.

Our proposal of the intermediate state with ordered magnetic charges corresponds to scenario (B). The charge-ordering transition is of the Ising type. It takes the system from the paramagnetic phase into a state where the Ising order parameter $\langle M^3 \rangle$ has a nonzero expectation value. If $\langle M^3 \rangle = \langle |M|^3 \re^{3\ri \phi}\rangle > 0$, the state can be thought of as a mixture of the clock states with $\phi = 0$, $2\pi/3$, and $-2\pi/3$, Fig.~\ref{fig:clock}(a). At a lower temperature, the system selects one of these states in a transition of the 3-state Potts universality class. 

To test which of the scenarios are realized in kagome spin ice, we have performed large-scale numerical simulations of both the short and long-range models with the Hamiltonians $H_1 + H_2$ and $H_1 + H_\mathrm{d}$.

\section{Numerical simulations}
\subsection{Kagome ice with short-range interactions}

\begin{figure}
\includegraphics[width=\columnwidth]{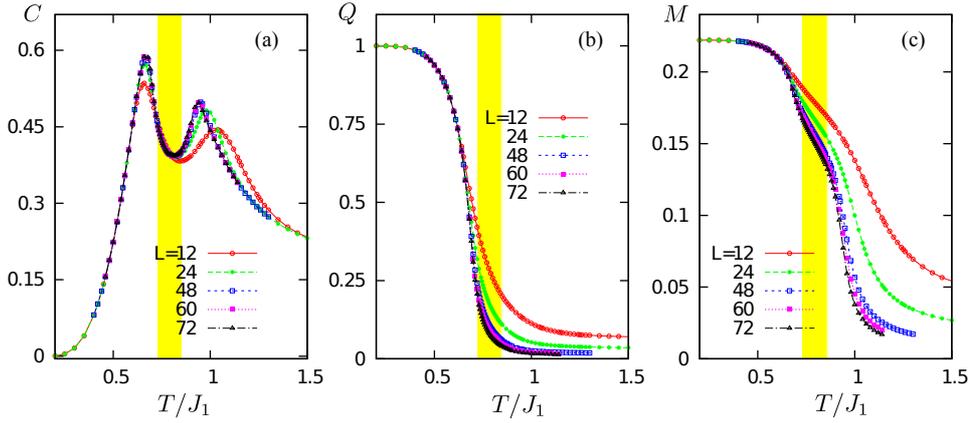}
\caption{Temperature dependence of (a) Specific heat $C = \left(\langle E^2\rangle - \langle E \rangle^2\right)/N 
k_B T^2$, (b) magnetic charge $Q$, and (c) magnetic order parameter $M$ for varying lattice sizes obtained from 
Monte Carlo simulations of the short-range spin ice model. The simulation was done
with $J_2 = -J_1/3$.  The shaded area indicates the intermediate critical phase. 
The transition temperatures $T_{\mathrm c1} = 0.735J_1$ and $T_{\mathrm c2} = 0.845J_1$ are determined using the 
finite-size scaling relation Eqs.~(\ref{eq:kt-scaling2}).}
\label{fig:j2-1}
\end{figure}

\begin{figure}
\center
\includegraphics[width=0.7\columnwidth]{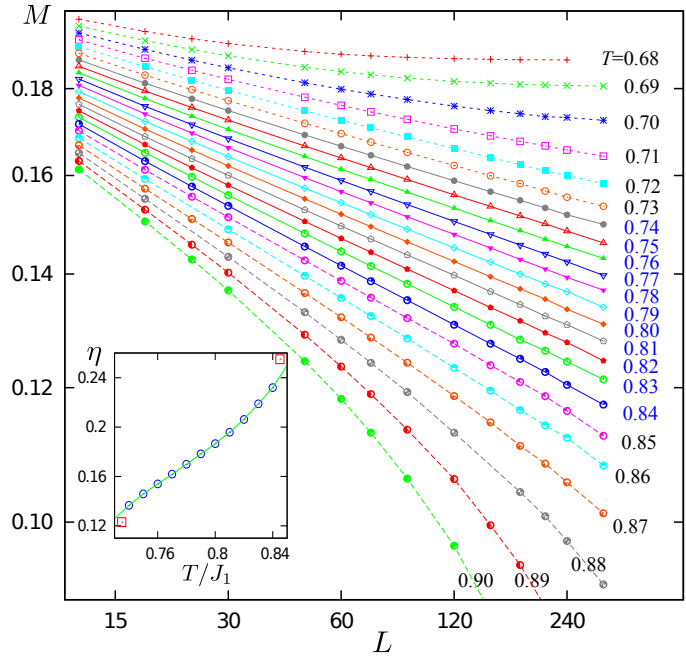}
\caption{The log-log plots of order parameter $M$ as a function of system size $L$ at various temperatures.  
The solid curves indicate the linear behavior which corresponds to a power-law dependence, $M \propto L^{-\eta/2}$,
in the intermediate critical phase. The inset shows the extracted critical exponent $\eta$ as a function
of temperature in the critical regime. The two data points represented by symbol $\boxdot$ are obtained
using the finite-size scaling Eqs.~(\ref{eq:kt-scaling2}).}
\label{fig:j2-2}
\end{figure}

Here we provide numerical evidence for scenario (A), namely two sucessive KT transitions, in the short-range 
ice model on kagome. A standard Metropolis importance sampling was used to simulate the behavior
of the model on $L\times L$ unit cells with periodic boundary conditions; the number of spins $N = 3L^2$.
In a single Monte Carlo step, each spin in the lattice is updated once with
a probability $p=\min(1, \exp(-\Delta E/k_B T))$, where $\Delta E$ is the energy change of flipping a  
spin. It is worth noting that the single-spin update is sufficient for both the spin-ice regime and even
the intermediate critical phase; the corresponding acceptance rates are roughly 30 \% and 15 \%, respectively.
About 10000 initial Monte Carlo sweeps were discarded to allow the system to equilibrate. We then retained
results from $10^5\sim 10^6$ Monte Carlo sweeps for computing averages. Our most complete data set was taken 
with parameters $J_2 = -J_1/3$, for which we describe our analysis in detail in the following.

Fig.~\ref{fig:j2-1} shows the temperature variation of specific heat, staggered charge order, and magnetic
order parameters for different lattice sizes. While the appearance of two peaks in the specific-heat 
curve clearly indicates two phase transitions, 
the locations and amplitudes of the peaks vary only slightly with lattice size, a feature characteristic
of KT transitions. The intermediate critical phase determined by finite-size scaling to be discussed
below is marked by the shaded region in the figures.
The charge order parameter is defined as difference of magnetic charges on the two different types of triangles:
\begin{eqnarray}
	Q = \frac{1}{2L^2} \sum_{\alpha} (-1)^\alpha\, Q_{\alpha},
\end{eqnarray}
where $(-1)^\alpha = \pm 1$ for up and down triangles, respectively, and $2L^2$ is the number of triangles
in the lattice. The charge order is proportional to the Ising order parameter, $\langle Q \rangle \propto 
\langle M^3 \rangle$. As can be seen from Fig.~\ref{fig:j2-1}(b), the staggered charge order decreases rapidly 
with increasing $L$ above the low-$T$ transition, indicating that magnetic charges remain disordered in the intermediate phase.

\begin{figure}
\includegraphics[width=\columnwidth]{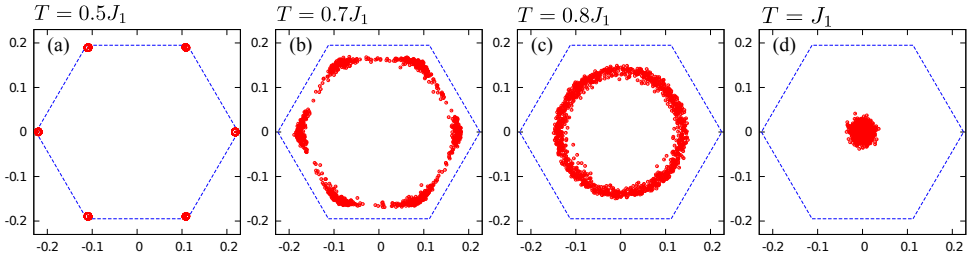}
\caption{Distribution of instantaneous magnetic order parameter $M$ in the complex plane $({\rm Re}M, {\rm Im}M)$
for different temperatures.}
\label{fig:j2-3}
\end{figure}

On the other hand, the magnetic order parameter shows quite different finite-size behavior across 
the three different regimes, Fig.~\ref{fig:j2-1}(c). Above the high-$T$ peak the rapid decrease of $M$ with
increasing $L$ is consistent with a magnetically disordered phase, whereas a negligible finite-size effect implies 
an ordered state below the second transition at $T_{\mathrm c1}$. In the intermediate regime the order parameter $M$ 
falls off rather slowly with increasing system size. These observations are summarized in Fig.~\ref{fig:j2-2} 
which shows magnetic order parameter $M$ as a function of $L$ at different temperatures. The magnetic order 
parameter extrapolates to zero at high temperatures, whereas it levels out to a constant at low $T$. 
In the intermediate regime, the linear behavior in the double-logarithmic plot indicates a power-law dependence
\begin{equation}
	\label{eq:M-scaling}
	M \propto L^{-\eta/2},
\end{equation}
which is consistent with a divergent correlation length in a critical phase. The extracted value of critical
exponent $\eta$ as a function of temperature is shown in the inset of Fig.~\ref{fig:j2-2}.

As discussed in Sec.~\ref{sec:model}(\ref{sec:transitions}), the sixfold anisotropy $a_6\cos6\phi$ 
in the Landau expansion of the free-energy functional is irrelevant in the intermediate 
critical phase. To confirm this, we show in Fig.~\ref{fig:j2-3} the distribution of instantaneous order 
parameter $M = |M| e^{{\rm i}\phi}$ in the complex $({\rm Re}M, {\rm Im}M)$ plane at various temperatures.
As can be seen in Fig.~\ref{fig:j2-3}(c) which corresponds to a temperature in the intermediate regime, 
the order parameter has a finite amplitude $|M| \neq 0$ due to finite system size ($L = 180$). 
More importantly, a continous O(2) symmetry emerges in this critical phase. In the ordered phase below $T_{\mathrm c1}$,
the anisotropy term reasserts itself and restores the $Z_6$ symmetry as demonstrated in 
Figs.~\ref{fig:j2-3}(a) and (b).

To further corroborate the proposed scenario of two KT transitions, we resort to the method of finite-size 
scaling (Challa \& Landau 1986). For a KT transition, the correlation length near the critical 
temperature $T_\mathrm{c}$ diverges as (Kosterlitz 1974)
\begin{equation}
	\xi \propto \exp (a\, t^{-1/2}),
\end{equation}
where $a$ is a non-universal constant and $t = |T - T_\mathrm{c}|/T_\mathrm{c}$ is the reduced temperature. 
The order parameter and its susceptibility exhibit power law behavior,
$M \propto \xi^{-\eta/2}$ and $\chi_\mathrm{M} \propto \xi^{2-\eta}$, in a KT transition.
For finite systems the singular part of the free energy is expected to depend only on $\xi/L$.
The order parameter and susceptibility are assumed to have the functional forms
\begin{equation}
	\label{eq:kt-scaling2}
	M = L^{-\eta/2}\mathcal{M}(\xi/L), \quad\quad \chi_\mathrm{M} = L^{2-\eta}\,\mathcal{X}(\xi/L),
\end{equation}
where $\mathcal{M}$ and $\mathcal{X}$ are unknown universal functions. At the critical point, 
substituting $\xi = \infty$ into Eq.~(\ref{eq:kt-scaling2}) gives the power law relation (\ref{eq:M-scaling}).
For an extended regime consisting of a line of critical points, the power law $M \sim L^{-\eta/2}$ should hold 
over the entire temperature range from $T_{\mathrm c1}$ to $T_{\mathrm c2}$, which is indeed observed in our simulation, Fig.~\ref{fig:j2-2}.

\begin{figure}
\center
\includegraphics[width=0.95\columnwidth]{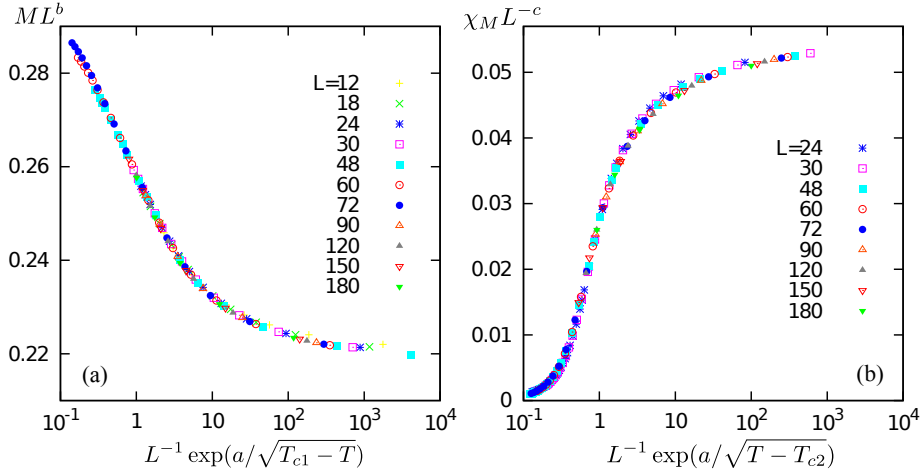}
\caption{(a) Finite-size scaling of the order parameter $M$ for the lower-temperature transition at $T_{\mathrm c1}$.
(b) Finite-size scaling of susceptibility $\chi_\mathrm{M} = (\langle M^2\rangle - \langle M\rangle^2)/N k_B T$ 
for the higher-temperature transition at $T_{\mathrm c2}$. 
The following values of the parameters are used: $a = 1.22$, $b = 0.0615$,
$c = 1.746$, $T_{\mathrm c1} = 0.735J_1$, and $T_{\mathrm c2} = 0.845J_1$.}
\label{fig:j2-4}
\end{figure}

The finite-size scaling relations~(\ref{eq:kt-scaling2}) can also be used to determine the upper and
lower critical temperatures of the intermediate KT phase. For example, with an appropriately
chosen set of parameters $a$, $b$, and $T_{\mathrm c1}$, the plot of $M L^b$ versus $L^{-1}\exp(a /\sqrt{T_{\mathrm c1}-T})$
should collapse on a universal curve for different system sizes. The same should hold for $\chi_\mathrm{M} L^{-c}$
versus $L^{-1}\exp(a /\sqrt{T-T_{\mathrm c2}})$. The two constants, $b$ and $c$, are related to the critical 
exponent: $b = \eta/2$ and $c = 2-\eta$. As shown in Fig.~\ref{fig:j2-4}, excellent data collapse was
obtained using the following set of parameters: $a = 1.22$, $b = 0.0615$, $c = 1.746$, $T_{\mathrm c1} = 0.735J_1$, 
and $T_{\mathrm c2} = 0.845J_1$. From this, we estimated the critical exponent at the two transition 
temperatures: $\eta(T_{\mathrm c1})=2b = 0.123$ and $\eta(T_{\mathrm c2}) =2-c = 0.255$.
These values agree resonably well with those extracted by linear-fitting of the power-law 
relation~(\ref{eq:M-scaling}) within the critical phase, see the inset of Fig.~\ref{fig:j2-2}.
Both values are close to the theoretical predictions $1/9$ and $1/4$ for the six-state clock model 
(Jos\'e \emph{et al.} 1977), and the slight overestimation could be due to the finite-size effects on KT transitions.

\subsection{Kagome ice with dipolar interactions}

\begin{figure}
\includegraphics[width=\columnwidth]{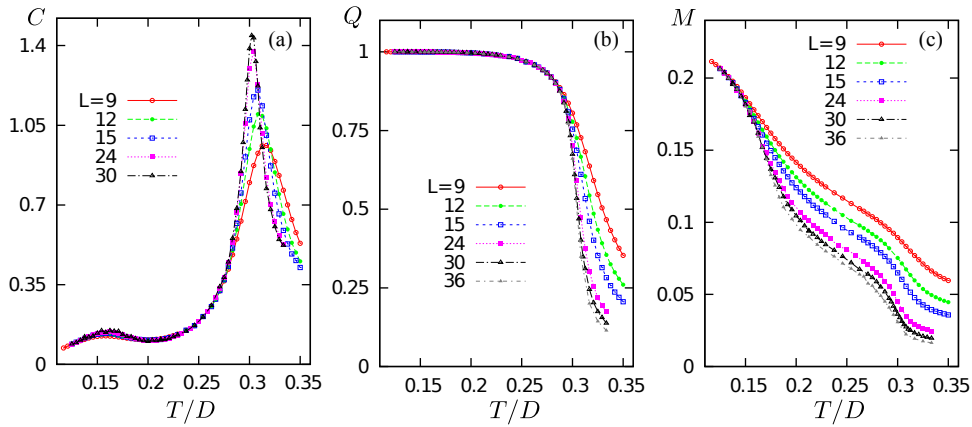}
\caption{Temperature dependence of (a) Specific heat $C = (\langle E^2\rangle - \langle E\rangle^2)/N k_B T^2$, 
(b) charge $Q$, and (c) magnetic $M$ order parameters for varying system sizes. 
The simulation was done with $D = 2J_1$.}
\label{fig:dip-1}
\end{figure} 

Although the single-spin Metropolis algorithm is quite efficient for simulating short-range kagome ice, 
it suffers from a dynamical freezing in the charge-ordered phase. A similar problem arises in the 
low-temperature simulation of pyrochlore spin ice (Melko \emph{et al.} 2001) in which the single-spin flip
violates the ice rules, leading to a large energy cost and low acceptance rate of the updates.
To overcome this problem we added the nonlocal loop moves first introduced by Barkema \& Newman (1998)
for square ice models. In a nonlocal update, a loop is first formed by randomly tracing a path through
triangles satisfying the ice rules, alternating between spins pointing in and spins pointing out of the
triangles. The process is completed when the path closes upon the starting spin or encounters any spin already
included in the loop. Flipping all the spins in such a loop leaves the
magnetic charges $\{Q_{\alpha}\}$ intact and conserves the nearest-neighbor energy $H_1$.
The loop move results in a small gain or loss of the dipolar energy $\Delta E_{\rm d}$, and the
update is accepted with a probability $p = \min(1, \exp(-\Delta E_{\rm d}/k_B T))$.

We employed a combination of single-spin flips and loop moves to simulate the long-range ice 
model $H_1 + H_{\rm d}$. With periodic boundary conditions, dipolar interactions were summed over
periodic copies up to a distance of $500L$. Most of the results presented here are obtained with a 
dipolar coupling $D = 2 J_1$. Fig.~\ref{fig:dip-1} shows the temperature dependence of specific heat, 
charge and magnetic order parameters for various system sizes.  Similar to the case of short-range 
kagome ice, two peaks can be seen in the specific-heat curves, indicating two phase transitions.
While the high-temperature peak becomes sharper with increasing $L$, the low-$T$ transition
barely shows any finite-size dependence. The behaviors of charge and magnetic order parameters shown
in Figs.~\ref{fig:dip-1}(b--c) are consistent with scenario~(B) discussed in the Introduction.
We discuss both transitions below.

\begin{figure}
\center
\includegraphics[width=0.9\columnwidth]{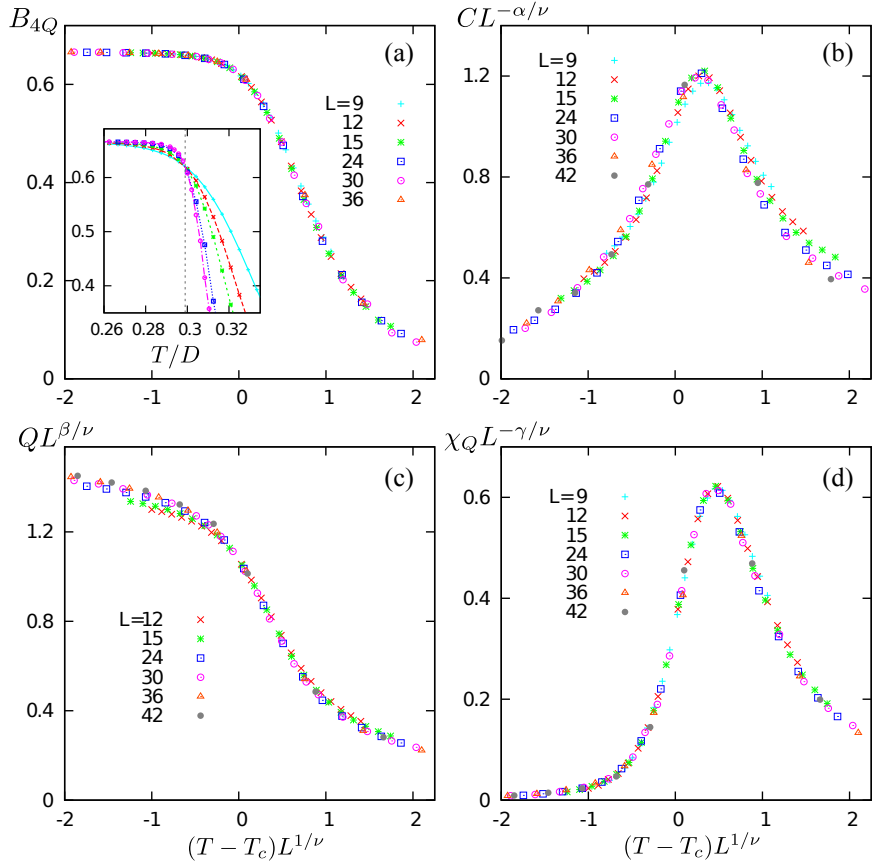}
\caption{Finite-size scaling of (a) Binder's fourth-order cumulant $B_{4Q} = 1-\langle Q^4 \rangle/3
\langle Q^2\rangle^2$, (b) specific heat, (c) staggered charge order parameter, and (d) charge susceptibility
$\chi_Q = (\langle Q^2\rangle - \langle Q \rangle^2)/N k_B T$ for the high-temperature transition.
The inset in (a) shows Binder's cumulant of different lattice sizes crossing at the critical temperature
$T_{\mathrm c2} = 0.29D$. The critical exponents of two-dimensional Ising model $\alpha = 0$, $\beta = 1/8$,
$\gamma = 7/4$, and $\nu = 1$ are used.}
\label{fig:dip-2}
\end{figure}

The staggered charge order corresponds to the partially ordered phase in scenario~(B) which is characterized by 
an Ising order parameter $\langle Q\rangle \propto \langle M^3 \rangle$ . 
The charge-ordering transition is thus expected to be in the universality class of the Ising model. To verify this
conjecture, we performed finite-size scaling analysis and found excellent data collapse, Fig.~\ref{fig:dip-2},
using the critical exponents of the two-dimensional Ising universality class. The critical temperature 
$T_{\mathrm c2} = 0.29D$ is determined from the crossing of the Binder's fourth-order cumulant for different lattice sizes.

The origin of an intermediate phase with ordered charges can be understood by expressing the energy of the
dipolar ice in terms of magnetic charges. This is achieved through the so-called dumbbell approximations 
(Castelnovo \emph{et al.} 2008) in which the dipoles are stretched into bar magnets of length $a = (2/\sqrt{3}) r_{\rm nn}$ such that their poles meet at the centers of triangles:
\begin{equation}
E(\{Q_\alpha\}) = \sum_{\alpha} \frac{v_0}{2} Q_\alpha^2
    + \frac{\mu_0 q_\mathrm{m}^2}{8\pi}\sum_{\alpha \neq \beta}
    \frac{Q_\alpha Q_\beta}{|\mathbf r_\alpha - \mathbf r_\beta|}.
\label{eq:E}
\end{equation}
Here $q_\mathrm{m} = \mu/a$ is the magnetic charge of the dumbbell, $\mu$ is the moment of the spin. The self-energy  
$v_0 = J/2 + (11+3\sqrt{3})D/8$ for kagome ice. It can be shown that the dipolar energy $H_1 + H_{\rm d}$ is
well approximated by Eq.~(\ref{eq:E}) with corrections which vanish with distance at least as fast as $1/r^5$ 
for each dipole pair.

\begin{figure}
\center
\includegraphics[width=0.85\columnwidth]{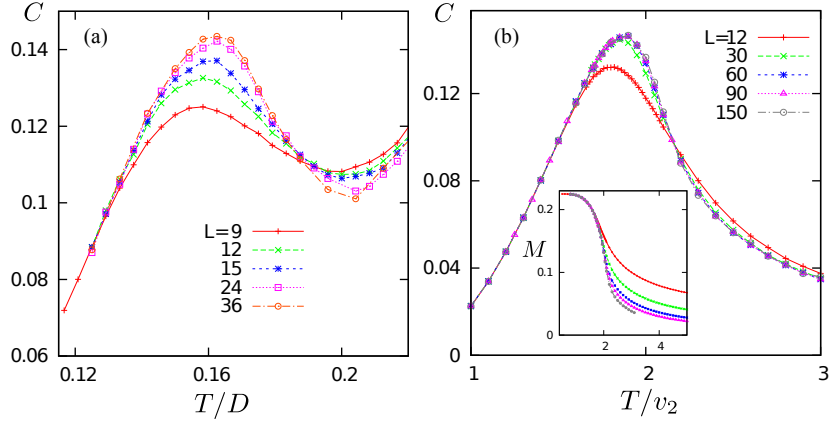}
\caption{Specific heat as a function of temperature for (a) dipolar ice model on kagome, and (b) dimer model
with next-nearest-neighbor attraction on the hexagonal lattice. The inset in (b) shows the temperature variation
of the order parameter characterizing the $\sqrt{3}\times\sqrt{3}$ dimer order.}
\label{fig:dip-3}
\end{figure}

The dumbbell approximation also illustrates an important difference bwteen pyrochlore and kagome
ices. Since the allowed charges are even on a tetrahedron and odd on a triangle, minimization of the self-energy 
term results in $Q_{\alpha} = 0$ on the pyrochlore lattice and $\pm 1$ on kagome. The conditions of minimal allowed charges on simplexes correspond to the ice rules on the respective lattices. For pyrochlore ice, the vanishing of
magnetic charge also makes the Coulomb interaction, the second term in Eq.~(\ref{eq:E}), strictly zero.
This observation explains why all states satisfying the ice rules are esstentially degenerate in pyrochlore 
spin ice over a wide range of temperatures (Gingras \& den Hertog 2001). The degeneracy is lifted by
the residual interactions neglected in the Coulomb approximation and a long-range magnetic order with
wavevector $\mathbf Q = (0, 0, 2\pi)$ sets in at a lower temperature compared to the dipolar energy scale
(Melko \emph{et al.} 2001).

The situation on kagome is quite different as nonzero charges generate a magnetic field.
This results in substantial energy differences between states obeying the ice rule $Q_{\alpha} = \pm 1$.
Consequently, interactions between uncompensated charges induce an Ising transition into a state with staggered 
charge order which minimizes the Coulomb energy. This partially ordered phase is closely related to the 
ice states of pyrochlore spin ice in a $\langle 111 \rangle$ magnetic field (Udagawa \emph{et al.} 2002; 
Moessner \& Sonhdi 2003). Triangles with positive (negative) charges on kagome correspond to 
three-in-one-out (three-out-one-in) tetrahedra of the pyrochlore lattice.

\begin{figure}
\includegraphics[width=1\columnwidth]{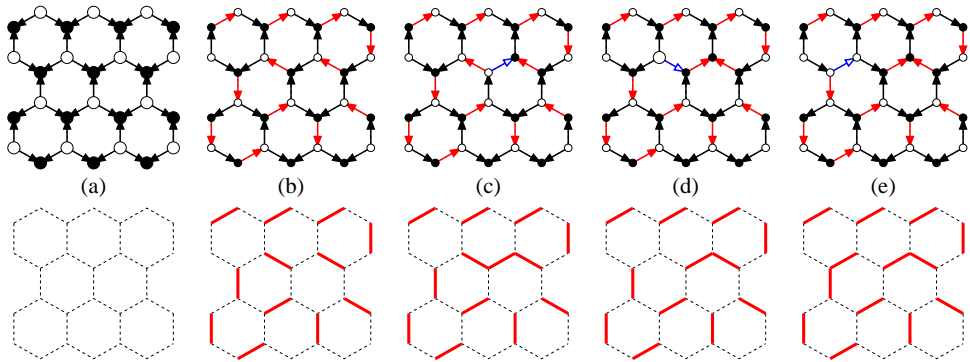}
\caption{Top panel: spin configurations. Bottom: respective dimer configurations. (a) Reference state with
charges $Q_\alpha=\pm 3$ on the two sublattices. (b) A charge-ordered state with $Q_\alpha = \pm 1$. (c--e)
two charge defects ($\Delta Q_{\alpha} = \pm 2$) are created and pulled apart. Red arrows denote the minority
spins.}
\label{fig:dimers}
\end{figure}

Spins remain disordered in the charge-ordered ice phase as evidenced by the loop moves discussed at the 
beginning of this section. Such nonlocal loop updates flip spins on a closed path while maintaining the charge
configuration. To quantify this residual degeneracy, we note that each triangle in a state with perfect 
charge order has two majority spins pointing into (or out of) the triangle and a minority spin pointing the 
other way.  Such states can be mapped to dimer coverings on the honeycomb lattice by identifying the minority 
spins with the dimers, Fig.~\ref{fig:dimers}(b). The number of dimer-coverings on honeycomb grows 
exponetially with the lattice size, giving rise to a residual entropy density $S = 0.108$ per spin 
(Udagawa \emph{et al.} 2002).

The remaining entropy of the charge-ordered phase is completely removed by a magnetic phase transition
corresponding to the low-$T$ peak of the specific-heat curve, Fig.~\ref{fig:dip-1}(a). The magnetic order
is shown in Fig.~\ref{fig:kagome}(d) and is again characterized by order parameter $M$; it has an enlarged 
$\sqrt{3}\times\sqrt{3}$ unit cell containing 9 spins.
Similar to pyrochlore spin ice, the magnetic ordering is induced by residual interactions beyond the dumbbell 
approximation Eq.~(\ref{eq:E}). 

As the $Z_6$ symmetry of the spin-ice Hamiltonian is reduced to a 
threefold symmetry in the charge-ordered phase, the magnetic transition is expected to be in the universality 
class of 3-state Potts model. However, Monte Carlo simulations on systems up to $L = 36$ fail to turn up any 
signature of the Potts criticality. Instead, the lack of a singularity in the specific heat, 
Fig.~\ref{fig:dip-3}(a), seems to be consistent with a KT transition. 

\subsection{Dimers on a honeycomb lattice}

To shed light on this puzzling result 
we consider a similar transition in the honeycomb dimer model. As discussed above, the charge-ordered states
can be uniquely mapped to dimer coverings on honeycomb. To induce the corresponding 
$\sqrt{3}\times\sqrt{3}$ ordering of dimers, we introduce a next-nearest-neighbor attractive interaction between 
the dimers. Note that nearest-neighbor dimers are precluded by the hardcore constraints that each lattice site 
is attached to exactly one dimer. The partition function of the modified dimer model reads
\begin{equation}
	\label{eq:Z}
	Z = \sum_{\mathcal{C}} \exp\left(\frac{v_2 N_{2}(\mathcal{C})}{k_B T}\right),
\end{equation}
where the sum is over all dimer coverings of the hexagonal lattice and $N_2(\mathcal{C})$ counts the number 
of next-nearest-neighbor pairs in a given covering $\mathcal{C}$. An attractive interaction between next-nearest-neighbor dimers corresponds to $v_2 > 0$.

We employed the direct-loop Monte~Carlo algorithm (Sandvik \& Moessner 2006) to simulate the dimer 
model~(\ref{eq:Z}). The nonlocal update is performed by initially breaking up an arbitrary dimer into a
pair of monomers and then moving one monomer across the lattice by flipping a sequence of dimers along a path.
The process is completed and a new dimer configuration is created once the two monomers meet and recombine 
with each other. Since detailed balance is maintained at each step of the monomer's movement, large number of 
dimers can be efficiently updated without rejection. This method significantly 
reduces the autocorrelation time in the Monte~Carlo process and allows us to simulate large lattices up to $L=150$.
Fig.~\ref{fig:dip-3}(b) shows the specific heat versus temperature for various system sizes; also shown in the 
inset is the temperature dependence of the order parameter characterizing the $\sqrt{3}\times\sqrt{3}$ dimer 
order. The rather weak finite-size dependence and a lack of singularity in specific heat again shows that the 
dimer ordering is characterized by a KT transition similar to the case of dimer model with aligning 
interactions on the square lattice (Alet \emph{et al.} 2006).

\begin{figure}
\center
\includegraphics[width=1\columnwidth]{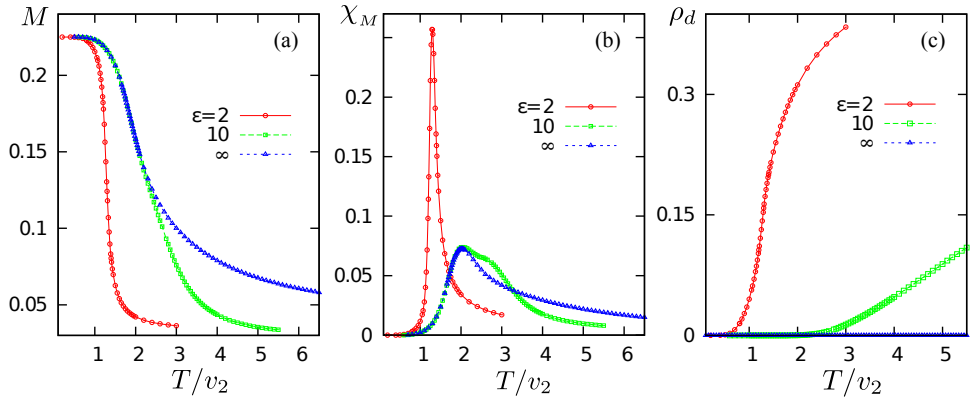}
\caption{Temperature variation of (a) order parameter $M$, (b) susceptibility $\chi_\mathrm{M} = (\langle M^2\rangle - 
\langle M \rangle^2)/N k_B T$, and (c) density of dimer-pair (charge defect) for phase transitions to a
long-range $\sqrt{3}\times\sqrt{3}$ dimer order. The three curves correspond to different fugacities
$z = \exp(-\varepsilon/k_B T)$ for the charge defects. The value $\varepsilon = \infty$ corresponds to models
with hardcore dimer covering. The simulations were done on a honeycomb lattice with $L=12$.}
\label{fig:dip-4}
\end{figure}

The appearance of a KT transition indicates the critical nature of the disordered hardcore dimer phase, or
equivalently the charge-ordered ice phase. Indeed, by further mapping the dimer covering to a `height' field
(Bl{\"o}te \& Hilhorst 1982),
the dimer model is described by a sine-Gordon model in the coarse-grained approximation.
It is well known that the height model undergoes a KT transition into a rough phase at high temperatures 
(Kardar 2007). As the confining cosine potential term is irrelevant at high-$T$, the rough phase is described 
by a Gaussian field theory with critical correlation functions in two dimensions.

\begin{figure}
\center
\includegraphics[width=0.9\columnwidth]{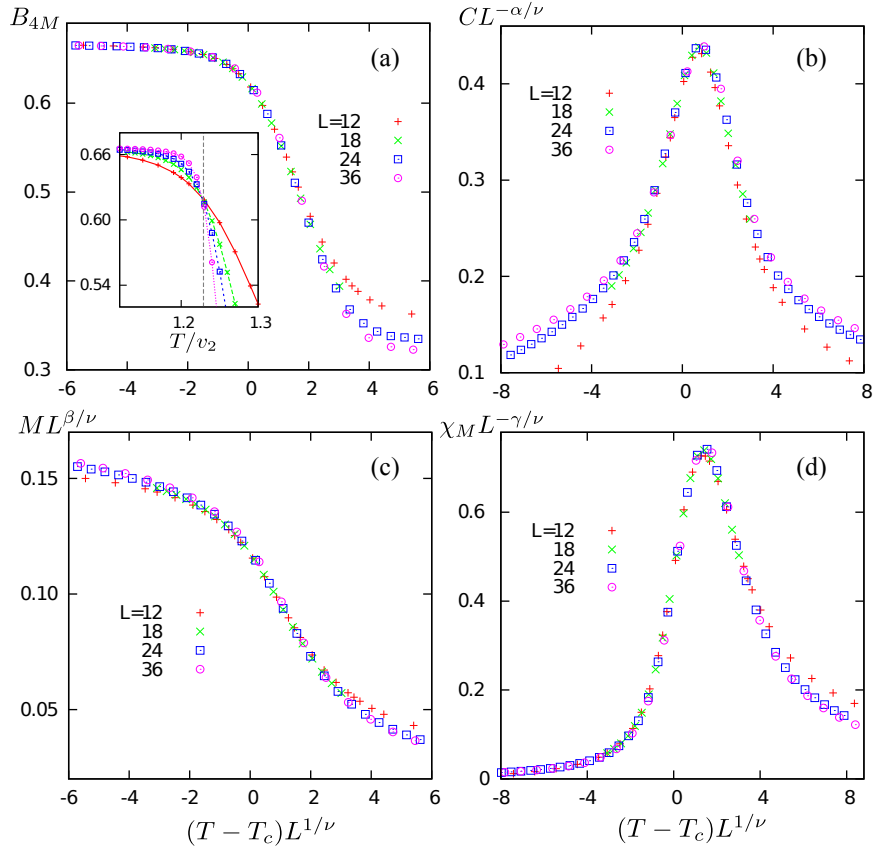}
\caption{Finite-size scaling of (a) Binder's fourth-order cumulant $B_{4M} = 1-\langle M^4 \rangle/3
\langle M^2\rangle^2$, (b) specific heat, (c) order parameter $M$, and (d) susceptibility 
$\chi_\mathrm{M} = (\langle M^2\rangle - \langle M \rangle^2)/N k_B T$ for the transition into the 
$\sqrt{3}\times\sqrt{3}$ dimer order. The inset in (a) shows Binder's cumulant of different lattice 
sizes crossing at the critical temperature $T_{M} = 1.226 v_2$. The critical exponents of two-dimensional 
3-state Potts model $\alpha = 1/3$, $\beta = 1/9$, $\gamma = 13/9$, and $\nu = 5/6$ are used.}
\label{fig:dip-5}
\end{figure}

The existence of such a critical phase relies on the absence of charge defects. At finite temperatures, however,
thermally excited defects spoil the mapping to hardcore dimer coverings. There are two types of defects in the 
charge-ordered phase: triangles with $\pm 3$ charges (ice defects) and triangles that violate the charge
order (charge defects). In the dimer model, the triply charged sites become monomers, whereas defects
in charge order become sites with two dimers, Fig.~\ref{fig:dimers}. We ignore the triply charged defects in the intermediate phase on account of their high energy cost and focus on the charge defects. To that end, we modified 
the honeycomb model~(\ref{eq:Z}) by allowing dimer pairs with fugacity $z = \exp(-\varepsilon/k_B T)$. The hardcore-dimer model is recovered in the limit $\varepsilon \to \infty$. Fig.~\ref{fig:dip-4} shows the 
temperature variation of the $\sqrt{3}\times\sqrt{3}$ order parameter, susceptibility and density of dimer 
pairs for $\varepsilon = 2$, $10$ and $\infty$ in units of $v_2$. The model with larger fugacity for dimer 
pairs shows a dramatically different behavior from that of hardcore dimer covering.

For $\varepsilon < +\infty$, a finite density of dimer pairs sets a bound on the dimer correlation length, thus
altering the criticality of magnetic ordering to the universality class of 3-state Potts model. 
To confirm this, we performed a finite-size scaling study on the dimer model with $\varepsilon = 2 v_2$. 
As can be seen from Fig.~\ref{fig:dip-4}, one obtains excellent data collapse with the critical exponents of the 
two-dimensional 3-state Potts model. However, when the average separation
between charge defects exceeds the lattice size, we are back to hardcore dimers with power-law spatial correlations for all distances. This explains the KT-like behavior observed in spin ice at 
small lattice sizes. The critical behavior characteristic of the 3-state Potts universality only reveals itself 
for sufficiently large systems (Otsuka 2011).

\section{Discussion}

Magnetic charges in spin ice are an example of an emergent phenomenon. Although the fundamental degrees of freedom in these materials are magnetic dipoles, the behavior of the system at low energies is often more conveniently expressed in the language of magnetic charges. In spin ice on kagome, magnetic charge of a simplex can take on odd values $\pm 1$ and $\pm 3$ in natural units. Even though magnetostatic interactions tend to minimize magnetic charge, low-energy spin-ice states have non-vanishing magnetic charges on simplexes. Such a magnet may have, in addition to the paramagnetic and fully ordered states, a distinct intermediate phase with ordered magnetic charges and disordered spins. This phase possesses considerable residual entropy (0.108 per spin) because a given pattern of magnetic charges can be realized by many configurations of magnetic dipoles. 

Our numerical simulations provide evidence for an intermediate phase with staggered magnetic charges in spin ice with dipolar interactions on kagome. As the system is cooled down from the paramagnetic state, it first gradually enters the spin-ice regime in which magnetic charges of triangles are restricted to the smallest (in magnitude) values of $\pm 1$. A phase transition of the Ising universality class takes the magnet into the intermediate phase with staggered magnetic charges. 

At a lower temperature, the system enters a magnetically ordered phase with six ground states shown in Figure~\ref{fig:clock}(a). Because only three of them are accessible from each of the two states of the charge-ordered phase, this transition is expected to be in the universality class of the 3-state Potts ferromagnet. However, observing the corresponding critical behavior proved challenging because the transition takes place in the background of nearly-perfect magnetic charge order. When the charge order has no defects at all, magnetic configurations can be mapped onto states of hard-core dimers with attraction. This mapping establishes that spins in the background of perfectly ordered magnetic charges are in a quasi-ordered phase with algebraic spatial correlations. The transition to the fully ordered phase is then of the Kosterlitz-Thouless type. Thermal fluctuations generate defects in charge order, thus violating the hard-core constraint for dimers. The spin correlation length is then set by the typical distance between isolated charge defects. As Figure~\ref{fig:dip-1}(b) shows, the charge order parameter quickly approaches saturation upon cooling below $T_{\mathrm c2} = 0.29 D$. At $T = 0.24 D$, i.e., considerably above the onset of magnetic order at $T_{\mathrm c1}$, the concentration of charge defects is about 0.01 per triangle.  Furthermore, most of these defects come in pairs with zero net charge and minimal separation (resulting from single spin flips), so they just renormalize the spin correlation function without shortening the correlation length. Not surprisingly, the intermediate phase exhibits power-law spin correlations up to very large distances and the transition to the magnetically ordered phase appears to be of the Kosterlitz-Thouless type. Observing the true critical behavior of the 3-state Potts universality class would require extremely large system sizes inaccessible to us. 

\begin{acknowledgements}

The authors thank Paula Mellado, Gunnar M\"oller, and Roderich Moessner for useful discussions. GWC was supported by  the U.S. National Science Foundation Grant No. DMR-0844115. OT was supported in part 
by the U.S. Department of Energy, Office of Basic Energy Sciences, Division of Materials Sciences and Engineering under Award No. DE-FG02-08ER46544. 
\end{acknowledgements}

\end{document}